\documentclass[onecolumn,authoryear]{els-mrw} 

\usepackage{amsmath,amssymb,amsfonts,amsthm,makeidx,graphicx}
\usepackage{txfonts}
\usepackage{helvet}

\newtheorem{ass}[theorem]{Assumption}

\makeatletter
\DeclareOldFontCommand{\rm}{\normalfont\rmfamily}{\mathrm}
\DeclareOldFontCommand{\sf}{\normalfont\sffamily}{\mathsf}
\DeclareOldFontCommand{\tt}{\normalfont\ttfamily}{\mathtt}
\DeclareOldFontCommand{\bf}{\normalfont\bfseries}{\mathbf}
\DeclareOldFontCommand{\it}{\normalfont\itshape}{\mathit}
\DeclareOldFontCommand{\sl}{\normalfont\slshape}{\@nomath\sl}
\DeclareOldFontCommand{\sc}{\normalfont\scshape}{\@nomath\sc}
\makeatother

\usepackage{csquotes}  

\usepackage{physics}   

\newcommand{\tm}{\times}

\newcommand \diag  {\operatorname{diag}}

\newcommand \eps {\varepsilon}

\newcommand \R   {\mathbb{R}}

\newcommand \A   {\mathcal{A}}

\newcommand \B   {\mathcal{B}}
\newcommand \K   {\mathcal{K}}
\newcommand \Kinf{\mathcal{K_\infty}}

\newcommand \KL  {\mathcal{KL}}

\newcommand \LL  {\mathcal{L}}

\newcommand \PD   {\mathcal{P}}

\newcommand{\Uc}{\ensuremath{\mathcal{U}}}

\newcommand{\vertiii}[1]{{\left\vert\kern-0.25ex\left\vert\kern-0.25ex\left\vert #1 
    \right\vert\kern-0.25ex\right\vert\kern-0.25ex\right\vert}}

\newcommand \qrq   {\quad\Rightarrow\quad}

\newcommand \qiq   {\quad\Iff\quad}

\newcommand \Iff   {\Leftrightarrow}

\newcommand \id  {\operatorname{id}}

\newcommand{\normt}[1]{{\left\vert\kern-0.25ex\left\vert\kern-0.25ex\left\vert #1 
		\right\vert\kern-0.25ex\right\vert\kern-0.25ex\right\vert}}

\newif\ifMath					
\newif\ifEngi					

\newif\ifDFGtext					 

\newif\ifAndo              
													
\newif\ifExercises					
\newif\ifSolutions          
\newif\ifGerman							
\newif\ifEnglish						

\newif\ifnothabil						

\newif\ifFuture							

\newif\ifConf                    
\newif\ifJournal								 

\newif\ifNOTFORBOOK
\newif\ifFullVersion
\newif\ifExludedDueToSpaceReasons

\usepackage{xifthen}

\newcommand{\einsnorm}[2]{\ensuremath{
    \!\!\;\!\!\!\;
    \left\bracevert\!\!\!\!\!\left\bracevert
    \!
		\ifthenelse{\isempty{#2}}{#1}{#1(#2)}
    \!
      \right\bracevert\!\!\!\!\!\right\bracevert
    \!\!\;\!\!\!\;
  }}

\usepackage{xcolor}
\definecolor{blond}{rgb}{0.98, 0.94, 0.75}
	
\newlength\mytemplen
\newsavebox\mytempbox

\makeatletter
\newcommand\mybluebox{%
    \@ifnextchar[
       {\@mybluebox}%
       {\@mybluebox[0pt]}}

\def\@mybluebox[#1]{%
    \@ifnextchar[
       {\@@mybluebox[#1]}%
       {\@@mybluebox[#1][0pt]}}

\def\@@mybluebox[#1][#2]#3{
    \sbox\mytempbox{#3}%
    \mytemplen\ht\mytempbox
    \advance\mytemplen #1\relax
    \ht\mytempbox\mytemplen
    \mytemplen\dp\mytempbox
    \advance\mytemplen #2\relax
    \dp\mytempbox\mytemplen
    \colorbox{blond}{\hspace{1em}\usebox{\mytempbox}\hspace{1em}}}

\makeatother

\makeatletter
\let\origd=\d
\renewcommand*\d{
  \relax\ifmmode
    \mathrm{d}%
  \else
    \expandafter\origd
  \fi
}\makeatother

\usepackage{mathtools}

\makeatletter 
\newcommand{\pushright}[1]{\ifmeasuring@#1\else\omit\hfill$\displaystyle#1$\fi\ignorespaces}
\newcommand{\pushleft}[1]{\ifmeasuring@#1\else\omit$\displaystyle#1$\hfill\fi\ignorespaces}
\makeatother 

\newcounter{syscounter}

\newcounter{WPcounter}
\newcounter{PRcounter}

\usepackage{pgfplots}
\usepackage{pgf}
\usepackage{tikz} 
\usetikzlibrary{decorations.pathmorphing} 
\usepackage{tikz-3dplot}
\usepgfplotslibrary{fillbetween}
\usetikzlibrary{arrows}
\usetikzlibrary{graphs,decorations.pathmorphing,decorations.markings}
\usetikzlibrary{calc,math}
\usetikzlibrary{patterns}
\usetikzlibrary{shapes}
\usetikzlibrary{tikzmark}

\usetikzlibrary {positioning}
\usetikzlibrary{shadows}
\usetikzlibrary{patterns.meta}
\usetikzlibrary{plotmarks}

\usepackage{subcaption}  

\usepackage[absolute,overlay]{textpos}
\usepackage{scalefnt}   

\tikzdeclarepattern{
  name=hatch,
  parameters={\hatchsize,\hatchangle,\hatchlinewidth},
  bounding box={(-.1pt,-.1pt) and (\hatchsize+.1pt,\hatchsize+.1pt)},
  tile size={(\hatchsize,\hatchsize)},
  tile transformation={rotate=\hatchangle},
  defaults={
    hatch size/.store in=\hatchsize,hatch size=5pt,
    hatch angle/.store in=\hatchangle,hatch angle=0,
    hatch linewidth/.store in=\hatchlinewidth,hatch linewidth=.4pt,
  },
  code={
      \draw[line width=\hatchlinewidth] (0,0) -- (\hatchsize,\hatchsize);
  }
}

\pgfmathsetmacro\weight{1/2}
\pgfmathsetmacro\third{1/3}
\pgfmathsetmacro\twothirds{2/3}

\tikzset{degil/.style={
            decoration={markings,
            mark= at position 0.5 with {
                  \node[transform shape] (tempnode) {$/$};
                  }
              },
              postaction={decorate}
}
}

\usetikzlibrary{shadows}
\tikzset{
diagonal fill/.style 2 args={fill=#2, path picture={
\fill[#1, sharp corners] (path picture bounding box.south west) -|
                         (path picture bounding box.north east) -- cycle;}},
reversed diagonal fill/.style 2 args={fill=#2, path picture={
\fill[#1, sharp corners] (path picture bounding box.north west) |- 
                         (path picture bounding box.south east) -- cycle;}}
}

\usepgfplotslibrary{fillbetween}
\usetikzlibrary{intersections,through}

\makeatletter 
\tikzset{use path/.code=\tikz@addmode{\pgfsyssoftpath@setcurrentpath#1}}
\makeatother

\definecolor{manipulator-color}{RGB}{88,44,44}
\definecolor{manipulator-contour}{rgb}{0.0, 0.18, 0.39}  

\tikzset{>=latex} 

\begin{document}

\chapter{Input-to-state stability of infinite-dimensional systems: Foundations and present-day developments}
\label{chap:ISS-for-PDEs}

\author[1]{Andrii Mironchenko}%
\author[2]{Christophe Prieur}%


\address[1]{\orgname{University of Klagenfurt}, \orgdiv{Department of Mathematics}, \orgaddress{9020, Klagenfurt, Austria}}
\address[2]{\orgname{Univ. Grenoble Alpes}, \orgdiv{CNRS, Grenoble-INP, GIPSA-lab}, \orgaddress{F-38000, Grenoble, France}}

\articletag{Chapter Article tagline: update of previous edition,, reprint..}

\maketitle

%
%
%

\begin{abstract}[Abstract]
Input-to-state stability (ISS) unifies the stability and robustness in one notion, and serves as a basis for broad areas of nonlinear control theory. 
In this contribution, we covered the most fundamental facts in the infinite-dimensional ISS theory with a stress on Lyapunov methods. 
We consider various applications given by different classes of infinite-dimensional systems. Finally, we discuss a Lyapunov-based small-gain theorem for stability analysis of an interconnection of two ISS systems.
\end{abstract}

\begin{glossary}[Keywords]
Distributed parameter systems; Stability; Robustness; Lyapunov functions
\end{glossary}

\begin{glossary}[Key points/objectives]
\begin{itemize}
\item Motivation to study the input-to-state stability of dynamical systems;
\item Discussion of direct and converse ISS Lyapunov theorems;
\item Focus is done on some applications using different partial differential equations;
\item Present the small-gain technique for  stability analysis of coupled systems.
\end{itemize}
\end{glossary}

\section{Introduction}


Modern applications of control theory to traffic networks \cite{yu2022traffic}, multi-phase systems \cite{koga2020materials}, fusion control \cite{AWP14}, and numerous other areas, require methods for stability analysis of coupled systems, described by partial differential equations (PDEs), and the analysis of the impact of the disturbances on the performance. The notion of input-to-state stability (ISS) allows us to study both these problems simultaneously, which underlines its crucial role in the robust stability analysis and control of nonlinear systems. ISS is instrumental in estimating the impact of disturbances and analyzing feedback interconnections of dynamical systems. This concept has been first introduced for finite-dimensional systems in \cite{Son89} and further developed for distributed parameters systems such as those described in abstract form or modeled by PDEs, see in particular \cite{MiP20}, \cite{Mir23}, and \cite{KaK19}. 
Nowadays ISS plays a key role in the development of the unifying framework for robust
stability analysis, robust control, and observation of infinite-dimensional nonlinear systems.

Lyapunov functions constitute one of the most fundamental tools to establish ISS yielding the characterization of ISS for a large class of infinite-dimensional systems, paralleling what is known for nonlinear finite-dimensional systems \cite{SoW95} (where an equivalence between ISS and existence of a smooth ISS Lyapunov function has been established). In this paper, we give an easy-to-access overview of the infinite-dimensional input-to-state stability theory.
We first introduce the general definition of ISS for systems in presence of disturbances and relate it to the stability and attractivity notions for systems with inputs. Then, results relying on direct Lyapunov functions are stated together with a converse Lyapunov result. 
We recall also some specific results proven in \cite{KaK19} for particular PDEs. 
For a detailed overview of ISS and for many more references, we refer to the survey \cite{MiP20}, and monographs \cite{KaK19,Mir23}. 

\textbf{Notation.} We denote $\R_+:=[0,+\infty)$.
For two sets $X,Y$ denote by $C(X,Y)$ the linear space of continuous functions, mapping $X$ to $Y$.
For the formulation of stability properties we use the standard classes of comparison functions:
\index{comparison!functions}
\index{class!$\K$}
\index{class!$\LL$}
\index{class!$\KL$}
\index{class!$\PD$}
\index{class!$\Kinf$}
\index{positive-definite function}
\index{function!positive definite}
\begin{equation*}
\begin{array}{ll}
{\K} &:= \left\{\gamma \in C(\R_+,\R_+) : \gamma(0)=0 \mbox{ and } \gamma \mbox{ is strictly increasing}   \right\},\\
{\K_{\infty}}&:=\left\{\gamma\in\K: \gamma\mbox{ is unbounded}\right\},\\
{\LL}&:=\{\gamma\in C(\R_+,\R_+): \gamma\mbox{ is strictly decreasing with } \lim\limits_{t\rightarrow\infty}\gamma(t)=0 \},\\
{\KL} &:= \left\{\beta \in C(\R_+\times\R_+,\R_+): \beta(\cdot,t)\in{\K},\ \forall t \geq 0, \  \beta(r,\cdot)\in {\LL},\ \forall r > 0\right\}.
\end{array}
\end{equation*}
An up-to-date compendium of results concerning comparison functions can be found in \cite[Appendix A]{Mir23} and \cite{Kel14}.

\section{General Definitions}
\label{sec:Systems-and-ISS-def}



Following \cite{MiP20}, we consider general infinite-dimensional control systems given by a triple $\Sigma=(X,\Uc,\phi)$, where 
$X$ is a normed vector space, called the \emph{state space}, endowed with the norm $\|\cdot\|$; 
$\Uc$ is a normed vector space, called the \emph{input space}, endowed with a norm still denoted with the same notation $\|\cdot\|$ when there is no ambiguity; and 
a map $\phi:D_{\phi} \to X$, $D_{\phi}\subseteq \R_+ \times X \times \Uc$ (called \emph{transition map}), such that for all $(x,u)\in X \tm \Uc$ it holds that $D_{\phi} \cap \big(\R_+ \times \{(x,u)\}\big) = [0,t_m)\tm \{(x,u)\} \subset D_{\phi}$, for a certain $t_m=t_m(x,u)\in (0,+\infty]$.
		
We assume that the input $u$ satisfies the axiom of shift invariance, and the axiom of concatenation. The transition map $\phi$ satisfies the identity, causality, and cocycle properties. Furthermore, we assume that $\phi$ is continuous in time. We refer to \cite{MiP20} for an extensive definition. 

There are several advantages in considering such abstract systems. On the one hand, ordinary and time-delay equations, as well as many classes of evolution partial differential equations (PDEs) belong to this general class of systems, which makes it possible to develop a truly overarching theory of distributed parameter systems. On the other hand, such approach allows us to decouple the well-posedness analysis from the stability analysis, and concentrate solely on the latter one.

\begin{definition}
\label{def:FC_Property} 
We say that a control system $\Sigma=(X,\Uc,\phi)$ is \emph{forward complete (FC)}, if 
for every $(x,u) \in X \times \Uc$ and for all $t \geq 0$ the value $\phi(t,x,u) \in X$ is well-defined. 
We say that $\Sigma$ has \emph{bounded reachability sets (BRS)}, if $\Sigma$ is forward complete  and for each $r>0$
\[
\sup_{\|x\|\leq r,\ \|u\| \leq r,\ t\in[0,r]}\|\phi(t,x,u)\| < \infty.
\]
\end{definition}

Forward completeness guarantees merely global existence of all solutions, and is in general weaker than BRS (see \cite[Example 2]{MiW18b}). In contrast, BRS property implies existence of uniform bounds on finite-time trajectories of families of solutions starting in bounded balls. This makes BRS a bridge between the pure well-posedness theory, which studies existence and uniqueness properties of solutions, and the
stability theory, which is interested in global in time bounds on families of solutions.



For simplicity, we consider only forward-complete control systems in this paper, if the contrary is not mentioned explicitly.

We proceed to the main concept for this note:
\begin{definition}
\label{def:ISS}
System $\Sigma=(X,\Uc,\phi)$ is called 
\begin{itemize}
	\item \emph{input-to-state stable
(ISS)}, if there exist $\beta \in \KL$ and $\gamma \in \Kinf$ 
such that for all $ x \in X$, $ u\in \Uc$ and $ t\geq 0$ it holds that
\begin {equation}
\label{iss_sum}
\| \phi(t,x,u) \| \leq \beta(\| x \|,t) + \gamma( \|u\|).
\end{equation}
	\item \emph{uniformly globally asymptotically stable at zero (0-UGAS)}, if there exists $\beta \in \KL$ 
such that for all $ x \in X$ and all $ t\geq 0$ it holds that
\begin {equation}
\label{eq:0UGAS}
\| \phi(t,x,0) \| \leq \beta(\| x \|,t).
\end{equation}
\end{itemize}
\end{definition}


  
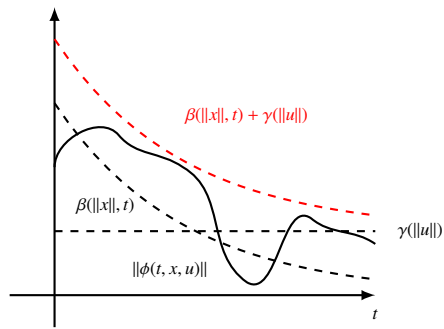
\begin{wrapfigure}{l}{0.4\textwidth}
    \centering
\begin{tikzpicture}[scale = 0.85, transform shape]
\draw[->,thick] (-0.7,0) --(5,0);
\draw[->,thick] (0,-0.5) --(0,4.5);
\draw[dashed,thick] (0,1) --(5,1);

\node(gain) at (5.7,1) {\small $\gamma(\|u\|)$};
\node(trans) at (0.8,1.4) {\small $\beta(\|x\|,t)$};
\node[red] (ISS_sum) at (3,2.8) {\small $\beta(\|x\|,t)+\gamma(\|u\|)$};

\node(x) at (1.8,0.4) {\small $\|\phi(t,x,u)\|$};
\node(t) at (5,-0.3) {\small $t$};

\draw[-,thick] (0,2) to [out=90,in=130](1,2.5) to [out=-50,in=145](2,2)
to [out=-30,in=145](3,0.2) to [out=-30,in=145](4,1.2)
to [out=-40,in=145](5,0.8);

\draw[dashed, thick,  domain=0:5]  plot (\x, {3*exp(-0.5*\x)});
\draw[dashed, thick, red, domain=0:5]  plot (\x, {1+3*exp(-0.5*\x)});
\end{tikzpicture}
    \caption{Typical solution of an ISS system with $u\not\equiv 0$}
		\label{fig:ISS-trajectories}
\end{wrapfigure}

A typical solution to an ISS system is depicted in Figure~\ref{fig:ISS-trajectories}.	
Substituting $u:=0$ into \eqref{iss_sum}, we see that ISS systems are necessarily 0-UGAS. 
In fact, one can show that solutions to $\Sigma$ corresponding to vanishing inputs ($\lim_{t \to \infty}\|u(\cdot +t)\| = 0$) converge to zero as $t\to\infty$.
Furthermore, for ISS systems, bounded inputs induce bounded state trajectories. 
In this way, ISS comprises both internal (or Lyapunov) and external (input-output) stability. 
Furthermore, the key tools from \emph{both} internal and external stability theories such as Lyapunov method including powerful converse Lyapunov theorems, superposition results, and small-gain theorems, have their counterparts in the ISS theory, and can be used in tandem.
This has made the ISS theory a cornerstone in the development of the nonlinear robust control theory.

For a normed linear space $W$ and any $r>0$ denote $B_{r,W} :=\{u \in W: \|u\| < r\}$ (the open ball of radius $r$ around $0$ in $W$).
If $W$ is the state space $X$, we write simply $B_r$ instead of $B_{r,X}$.

We are going to relate ISS to the stability and attractivity-like concepts stated next:
\begin{definition}
\label{def:ULS} 
$\Sigma$ is \emph{uniformly locally stable (ULS)}, that is, there exist $\sigma,\gamma \in\Kinf$ and $r>0$ such that whenever $\|x\|\leq r$,\ $\|u\| \leq r$,\ and $t \geq 0$, we have
\begin{equation}
\label{GSAbschaetzung}
  \left\| \phi(t,x,u) \right\| \leq \sigma(\|x\|) +\gamma(\|u\|).
\end{equation}
\end{definition}

ULS is a natural extension of local Lyapunov stability used in the dynamical systems theory to the systems with inputs.

\begin{definition}
\label{def:Limit-properties}
We say that a forward complete system $\Sigma$ has \emph{uniform limit (ULIM) property}, if there is
    $\gamma\in\Kinf$ so that for every $\eps>0$ and for every $r>0$ there
    exists a $\tau = \tau(\eps,r)$ such that 
for all $x\in B_r$ and all $u\in B_{r,\Uc}$ there is a $t\leq\tau$ such that 
\begin{eqnarray}
\|\phi(t,x,u)\| \leq \eps + \gamma(\|u\|).
\label{eq:sLIM_ISS_section}
\end{eqnarray}
\end{definition}

Trajectories of systems possessing the ULIM property, starting in a ball $B_r$ and subject to inputs of a magnitude uniformly bounded by $r$, intersect the $\varepsilon$-neighborhood of the ball $B_{\gamma(\|u\|_{\Uc}),X}$ within the interval $[0,\tau(\varepsilon,r)]$. After intersecting this neighborhood, the trajectories may well leave it, but then they have to visit it again and again. 
Thus, ULIM property can be understood as an extension of the weak attractivity concept to the systems with inputs.
The following result that relates ISS to stability and attractivity of systems with inputs, plays a central role in the ISS theory. 

\begin{theorem}[ISS superposition theorem, \cite{MiW18b}]
\label{thm:ISS-superposition} 
$\Sigma$ is ISS if and only if $\Sigma$ is BRS $\wedge$ ULS $\wedge$ ULIM.
\end{theorem}

The ISS superposition theorem is seldom used for direct verification of ISS for specific control systems, but it is used as a metatool helping to develop further tools for ISS analysis such as non-coercive ISS Lyapunov theorems and small-gain results.

\section{General techniques for ISS}

\subsection{Direct Lyapunov results}

Lyapunov theory is indispensable for ISS analysis of nonlinear infinite-dimensional systems. 
First of all, construction of an ISS Lyapunov function is in most cases the only realistic way to prove ISS of a given control system. 
Besides serving as certificates for the input-to-state stability, Lyapunov functions can be used for control purposes, such as robust stabilization, event-based control, ISS feedback redesign, etc., see \cite[Chapter 5]{Mir23}. Furthermore, having constructed ISS Lyapunov functions for subsystems of a finite or infinite network of control systems, one can use small-gain theorems to show (under certain conditions) the ISS of the whole network.
In this section, we introduce the basics of Lyapunov theory for distributed parameter systems.

For a real-valued function $b:\R_+\to\R$, define the \emph{right-hand upper Dini derivative} at $t\in\R_+$ by
\begin{eqnarray*}
D^+b(t) := \mathop{\overline{\lim}} \limits_{h \rightarrow +0} \frac{b(t+h) - b(t)}{h}.
\end{eqnarray*}

Let $x \in X$ and $V$ be a real-valued function defined in a neighborhood of $x$. \emph{The Lie derivative 
of $V$ at $x$ corresponding to the input $u$ along the corresponding trajectory of $\Sigma =(X,\Uc,\phi)$} is defined by
\begin{equation}
\label{ISS_LyapAbleitung}
\dot{V}_u(x):=D^+ V\big(\phi(\cdot,x,u)\big)\Big|_{t=0}=\mathop{\overline{\lim}} \limits_{t \rightarrow +0} {\frac{1}{t}\big(V(\phi(t,x,u))-V(x)\big) }.
\end{equation}

We are now in position to introduce the main notion of this section.
\begin{definition}
\label{def:noncoercive_ISS_LF}
A continuous function $V:X \to \R_+$ is called a \emph{non-coercive ISS Lyapunov function} for the system $\Sigma = (X,\Uc,\phi)$,
 if there exist $\psi_2,\alpha \in \Kinf$, $\sigma \in \K$ such that 
\begin{equation}
\label{LyapFunk_1Eig_nc_ISS}
0 < V(x) \leq \psi_2(\|x\|), \quad \forall x \in X\backslash\{0\},
\end {equation}
and the Lie derivative of $V$ along the trajectories of $\Sigma$ satisfies
\begin{equation}
\label{DissipationIneq_nc}
\dot{V}_u(x) \leq -\alpha(\|x\|) + \sigma(\|u\|)
\end{equation}
for all $x \in X$ and $u\in \Uc$.

If additionally there is a $\psi_1\in\Kinf$ so that 
\begin{equation}
\label{LyapFunk_1Eig_LISS}
\psi_1(\|x\|) \leq V(x) \leq \psi_2(\|x\|), \quad \forall x \in X,
\end {equation}
then $V$ is called a \emph{(coercive) ISS Lyapunov function} for the system $\Sigma = (X,\Uc,\phi)$.
\end{definition}

The classic result, which plays a pivotal role in the applications of Lyapunov theorem is:
\begin{theorem}{\emph{(Direct coercive Lyapunov theorem)}}
    \label{t:ISSLyapunovtheorem:1}
Let $\Sigma$ be a forward complete control system.
If there exists a coercive ISS Lyapunov function for $\Sigma$, then $\Sigma$ is ISS.    
\end{theorem}

From the viewpoint of nonlinear finite-dimensional control theory, the coercivity of Lyapunov functions seems to be a natural requirement. 
However, for infinite-dimensional systems, this assumption may be too limiting. 
Even for linear systems \eqref{eq:Linear_System} without inputs, if a semigroup is exponentially stable, the classical quadratic Lyapunov functions constructed by solving the operator Lyapunov equation \cite[Theorem 4.1.3]{CuZ20}, are non-coercive. 
This motivates the use of non-coercive Lyapunov functions introduced and justified for nonlinear systems in \cite{MiW19a}.
\begin{theorem}{\emph{(Direct non-coercive Lyapunov theorem, \cite[Theorem 3.7]{JMP20})}}
    \label{t:ISSLyapunovtheorem}
Let $\Sigma$ be a forward complete control system, which is CEP and BRS.    
If there exists a non-coercive ISS Lyapunov function for $\Sigma$, then $\Sigma$ is ISS.    
\end{theorem}

As discussed in \cite[Remark 2.15]{MiP20}, due to \eqref{ISS_LyapAbleitung}, the computation of the left-hand side of \eqref{DissipationIneq_nc} asks for the knowledge of the solution for small future times. For many classes of systems such as ODEs or delay systems with Lipschitz continuous right hand sides, one can find expressions for Lie derivative of $V$ that rely on the gradient of $V$ and on Driver's derivative of $V$ respectively. In many cases, Lie derivative can be computed based on Fr\'echet differentiability (see e.g., \cite[Definition A.5.31]{CuZ20}). 
To do that, we assume that the dynamics for $\Sigma$ are given in abstract form by 
\begin{equation}
\label{eq:dynamics}
\dot x= \mathcal{A}(x) +\mathcal{B}(u),
\end{equation}
where $\mathcal{A}$ is the generator on a strongly continuous (possibly nonlinear) semigroup and $\mathcal{B}$ is a (possibly nonlinear and unbounded) input operator. Well-posedness results for strong and mild solutions to (\ref{eq:dynamics}) are given, e.g., in\ \cite{TuW09,Sho13,miyadera1992nonlinear} for certain classes of nonlinear infinite-dimensional systems $\Sigma$.

If $V$ is Fr\'echet differentiable, then for systems \eqref{eq:dynamics} under certain assumptions on $\A, \B$, and $V$, one can reformulate the dissipativity condition \eqref{DissipationIneq_nc} as 
\begin{align}
\label{eq:Dissipative-inequality}
D V (x)( \mathcal{A}(x) +\mathcal{B}(u))\leq -\alpha (\|x\|)+\gamma (\|u\|),
\end{align}
which has to hold for all $x \in X$ and $u$ in $\mathcal{U}$ belonging to the domains of definitions of $\A$ and $\B$. 

This approach has been used in many works including \cite{zheng2018iss} on Burger's equation, \cite{KaK19} for parabolic and linear or semilinear hyperbolic PDEs, and \cite{bastin2021input} for quasilinear hyperbolic systems with boundary input disturbances. See also the seminal work \cite{PrM12} for the derivations of  sufficient conditions for dissipativity inequalities and thus for input-to-state stability properties of hyperbolic systems, relying on control Lyapunov functions as introduced in \cite{BaC16}.

Focusing on parabolic equations, Sturm-Liouville theory allows
to generalize what was done for finite-dimensional systems. Indeed, for a large class of reaction-diffusion equations, it is possible to compute a Riesz basis diagonalizing the dynamics of $\Sigma$ so that there exists a finite number of unstable modes. This allows to follow the machinery for nonlinear finite dimensional systems \cite{Mir23} and to generalize it for parabolic equations for various input-to-state operators, see, e.g., \cite{shreim2022input,lhachemi:prieur:book,PISANO2017447}.



\subsection{Converse Lyapunov theorem}

Direct Lyapunov theorems show that the existence of a Lyapunov function implies certain stability property. 
Converse ISS Lyapunov results claim that for an ISS system there is an ISS Lyapunov function of a certain type. 
The importance of such results is twofold: on the one hand, before we start looking for an ISS Lyapunov function of a given system, we would like to know that such a function exists at all. On the other hand, ISS Lyapunov functions can be used for control purposes, such as robust stabilization, event-based control, ISS feedback redesign, etc., see \cite[Chapter 5]{Mir23}. 
 

Consider infinite-dimensional systems of the form
\begin{equation}
\label{InfiniteDim}
\dot{x}(t)=Ax(t)+f(x(t),u(t)),
\end{equation}
where $A: D(A)\subset X \to X$ generates a $C_0$-semigroup $T(\cdot)$ of bounded linear operators, $X$ and $U$ are Banach spaces, and $f:X\times U \to X$ is Lipschitz continuous on bounded balls from the Banach space $(X\times U, \|\cdot\|_X+\|\cdot\|_U)$ to the
space $X$.

Taking the space $\Uc:=PC_b(\R_+,U)$ of globally bounded, piecewise continuous functions from $\R_+$ to $U$ which are right-continuous
as the space of admissible inputs for \eqref{InfiniteDim}, the system \eqref{InfiniteDim} becomes a well-posed (in the sense of mild solutions) control system.
For such systems we have
\begin{theorem}{\emph{(Converse ISS Lyapunov theorem, \cite[Theorem 5]{MiW17c})}}
\label{ISS_Converse_Lyapunov_Theorem}
System \eqref{InfiniteDim} is ISS if and only if there is a Lipschitz continuous on bounded balls coercive ISS Lyapunov function for \eqref{InfiniteDim}.
\end{theorem}

The fact that $f$ is Lipschitz continuous with respect to both variables is essentially used in the proof of Theorem~\ref{ISS_Converse_Lyapunov_Theorem}. In particular, the proof in \cite[Theorem 5]{MiW17c} does not work if $f$ is merely continuous w.r.t. $u$. 
Such assumptions are reasonable for PDEs with distributed inputs, but they are as a rule not satisfied for systems with boundary or point inputs. 
Existence of non-coercive ISS Lyapunov functions for some classes of linear systems with admissible operators has been shown in \cite{JMP20}.
Still, the derivation of converse ISS Lyapunov theorems for general linear systems with admissible input operators remains a challenging open problem.

\subsection{Semigroup and admissibility methods}

In this section, we present an important alternative to a Lyapunov method, which can be used for linear systems
\begin{equation}
\label{eq:Linear_System}
\dot{x}(t)=Ax(t)+ Bu(t), \quad \forall t \in\R_+,
\end{equation}
where $A$ is the generator of a strongly continuous semigroup $T:=(T(t))_{t\ge 0}$ on a Banach space $X$ and 
$B$ is unbounded as an operator from some Banach space $U$ to $X$, but $B$ is a bounded as an operator from $U$ to a larger space $X_{-1}$, that is, 
$B\in L(U,X_{-1})$, where we define the extrapolation space $X_{-1}$ as the completion of $X$ with respect to the norm 
$\|x\|_{X_{-1}}:= \|(aI -A)^{-1}x\|_X$ for some $a$ in a resolvent set of $A$.

Lifting of the state space $X$ to a larger space $X_{-1}$ brings several good news: the semigroup  $(T(t))_{t\ge 0}$ extends uniquely to a strongly continuous semigroup  $(T_{-1}(t))_{t\ge 0}$ on $X_{-1}$ whose generator $A_{-1}:X_{-1}\to X_{-1}$ is an extension of $A$ with $D(A_{-1}) = X$.

This allows to define mild solutions of \eqref{eq:Linear_System} for any $x \in X$ and any $u\in\Uc$ by means of the variation of constants formula:
\begin{eqnarray}
\phi(t,x,u)&:=&T_{-1}(t)x+\int_0^t T_{-1}(t-s)B u(s)ds\nonumber\\
        &=&T(t)x+\int_0^t T_{-1}(t-s)B u(s)ds.
\label{eq:Lifted_Lin_Sys_mild_Solution}
\end{eqnarray}
The last transition is due to the fact that $T_{-1}(t)x = T(t)x$ for all $x\in X$ and $t\geq 0$.

The lifting comes however at a price that now the solution $\phi(t,x,u)$ has values in $X_{-1}$.
The formula \eqref{eq:Lifted_Lin_Sys_mild_Solution} defines an $X$-valued function only in case if the value of the integral in \eqref{eq:Lifted_Lin_Sys_mild_Solution} belongs to the state space $X$, despite the fact that what we integrate is in $X_{-1}$.

This motivates the following definition: 
\begin{definition}
\label{def:q-admissibility} 
The operator $B\in L(U,X_{-1})$ is called a {\em $q$-admissible control operator} for $(T(t))_{t\ge 0}$, where $1\le q\le \infty$, if
there is a $t>0$ so that
\begin{eqnarray}
u\in L_q(\R_+,U) \qrq \int_0^t T_{-1}(t-s)Bu(s)ds\in X.
\label{eq:q-admissibility}
\end{eqnarray}
\end{definition}
One can see that a linear system \eqref{eq:Linear_System} with $\Uc:=L_q(\R_+,U)$, $q \in[1,\infty)$ is forward complete $\Iff$ \eqref{eq:Linear_System} is BRS $\Iff$ $B$ is $q$-admissible, see \cite[Propositions 3.10 and 3.11]{MiP20}.

An efficient criterion for ISS of linear systems is given by the following result shown in \cite[Proposition 2.10]{JNP18}:
\begin{theorem}
\label{thm:ISS-Criterion-lin-sys-with-unbounded-operators}
Let $X$ and $U$ be Banach spaces and let $\Uc:=L_p(\R_+,U)$ for some $p\in[1,+\infty]$. 
If $p=+\infty$, assume further that $\phi$ is continuous w.r.t. time in the norm of $X$.
Then 
\begin{center}
\eqref{eq:Linear_System} is ISS $\qiq$ $T$ generates an exponentially stable semigroup \ \ $\wedge$ \ \ $B$ is $p$-admissible.
\end{center}
\end{theorem}
Methods for analysis of the exponential stability of $C_0$-semigroups and the admissibility of unbounded input operators are well-developed \cite{Nee95,JaP04}. Theorem~\ref{thm:ISS-Criterion-lin-sys-with-unbounded-operators} allows us to use all this machinery to study ISS of linear distributed parameter systems.

\section[Interconnections of ISS systems]{Interconnections of input-to-state stable systems}
\label{GekoppelteISS_Systeme}

Stability analysis of coupled nonlinear systems is a complex task, especially if sub-systems are distributed parameter systems.
One of strengths of the ISS theory is that it gives such a powerful tool as nonlinear small-gain theorems, which allows to study stability of complex systems with ISS components.
As in most cases, the ISS of nonlinear systems is verified by the construction of an appropriate ISS Lyapunov function,
a natural desire is to use in the formulation of the small-gain theorems the information about the ISS Lyapunov functions for subsystems.
In this section, we present one of the most basic results of this type. 

Consider an interconnected system of the form
\begin{equation}
\label{Kopplung_N_Systeme}
\left\{
\begin{array}{l}
\dot{x}_{1}=A_{1}x_{1}+f_{1}(x_{1},x_{2},u), \\
\dot{x}_{2}=A_{2}x_{2}+f_{2}(x_{1},x_{2},u),
\end{array}
\right.
\end{equation}
where $x_1(t) \in X_1$, $x_2(t) \in X_2$, the spaces $X_1$ and $X_2$ are Banach spaces, and $A_i$ is the generator of a $C_{0}$-semigroup on $X_{i}$, $i=1,2$.
The input space we take as $\Uc:=PC_b(\R_+,U)$, i.e. as the space of piecewise continuous and right-continuous functions from $\R_+$ to some Banach space of input values $U$. 
The state space of the system \eqref{Kopplung_N_Systeme} we define as
\[
X=X_{1}\times X_{2},
\]
which is a Banach space with the norm 
\[
\|x\|
:=\|x_1\|
+\|x_2\|
,\quad \forall x=(x_1,x_2) \in X.
\]

The transition map of the $i$-th subsystem we denote by 
\[
\phi_i:\R_+ \times X_i \times (PC_b(\R_+, X_{3-i}) \tm \Uc) \to X_i. 
\]
For $x=(x_1,x_2) \in X$, define
\[
f(x,u):=(f_{1}(x,u)^\top,f_{2}(x,u)^\top)^\top,
\]
and
\[
A:=\diag(A_{1},A_{2}),
\quad
D(A):=D(A_{1}) \times D(A_{2}).
\]
Clearly, $A$ is the generator of a $C_{0}$-semigroup on $X$.
We rewrite the system \eqref{Kopplung_N_Systeme} in the vector form:
\begin{equation}
\label{KopplungHauptSys}
\dot{x}=Ax+f(x,u).
\end{equation}

We assume that 
\begin{ass}
\label{ass:Assumption-network} 
The system \eqref{KopplungHauptSys} satisfies the following properties:
\begin{enumerate}
	\item $f$ is continuous on $X \tm U$;
	\item For each $i=1,2$, $f_i$ is Lipschitz continuous w.r.t.\  $x_i$ on bounded subsets;
	\item Each subsystem is forward complete;
	\item (Well-posedness) For any initial condition $x \in X$ and any input $u\in\Uc$, there is a unique maximal solution of \eqref{KopplungHauptSys};
	\item If a maximal solution of \eqref{KopplungHauptSys} is uniformly bounded on its whole domain of existence, then it is defined on the whole $\R_+$ (boundedness-implies-continuation property).
\end{enumerate}
\end{ass}

Furthermore, we assume that both subsystems are ISS and
\begin{ass}
\label{ass_subsystem_iss}
For each $i=1,2$, there exists a continuous function $V_i:X_i \rightarrow \R_+$, that satisfies the following properties:
\begin{enumerate}
\item There exist $\psi_{i1},\psi_{i2} \in \Kinf$ such that%
\begin{equation}
\label{eq_subsystem_iss_coerc}
  \psi_{i1}(\|x_i\|
  ) \leq V_i(x_i) \leq \psi_{i2}(\|x_i\|
  ),\quad \forall x_i \in X_i;%
\end{equation}
\item There exist $\chi_{12},\chi_{21} \in \Kinf\cup\{0\}$, and $\chi_{1u}, \chi_{2u} \in \K$, $\alpha_1, \alpha_2 \in \PD$ such that for all $x_{j}\in X_j$, $j=1,2$, and all $u\in \Uc$ the following implications hold:%
\begin{align}
\label{eq_subsystem_orbitalder_est1}
\begin{split}
  V_1(x_1) > \max\Bigl\{\chi_{12}(x_2),\chi_{1u}(\|u\|
  ) \Bigr\} \Rightarrow \dot{V}_{1,u}(x_1) \leq - \alpha_1(V_1(x_1)),
\end{split}
\end{align}
\begin{align}
\label{eq_subsystem_orbitalder_est2}
\begin{split}
  V_2(x_2) > \max\Bigl\{\chi_{21}(x_1),\chi_{2u}(\|u\|
  ) \Bigr\} \Rightarrow \dot{V}_{2,u}(x_2) \leq - \alpha_2(V_2(x_2)).
\end{split}
\end{align}
\end{enumerate}
The functions $\chi_{12}, \chi_{21}$ are called \emph{internal Lyapunov gains}, while the functions $\chi_{1u},\chi_{2u}$ are called \emph{external Lyapunov gains}.%
\end{ass}
%
%
%
%

\begin{theorem}[Lyapunov-based ISS small-gain theorem]
\index{small-gain theorem!in Lyapunov formulation!for ODE systems}
\label{Constr_of_LF_FinDim} 
Let Assumptions above hold, and let the small-gain condition
\begin{eqnarray}
\chi_{12} \circ \chi_{21}(r) < r,\quad \forall r>0
\label{eq:SGT-2-sys}
\end{eqnarray}
be satisfied. Then \eqref{KopplungHauptSys}  is ISS.
%
\end{theorem}

A counterpart of Theorem~\ref{Constr_of_LF_FinDim} for arbitrary finite number of coupled systems has been shown in \cite{DaM13}, motivated by the corresponding small-gain results for interconnections of finite-dimensional systems in \cite{DRW10,JMW96}. 
For the analysis of interconnections of infinitely many systems, which is a topic actively studied during the last several years, we refer to \cite{KMZ23} (and references therein).

\subsection*{Example: coupling of nonlinear systems}

Consider now a coupling of two nonlinear parabolic systems:
\begin{equation}
\label{eq:Gekoppelte-parabolic-sys-ISS}
\left
\{
\begin {array} {l}
\partial_ t{x_{1}(z,t) =  q_1 \partial_ z^2x_{1}(z,t) + x^2_2(z,t), \quad \forall (z,t) \in (0,\pi)\times \R_+,}\\
{x_1(0,t) = x_1(\pi,t)=0, \quad \forall t \in  \R_+;} \\
\partial_ t{x_{2,t}(z) =  q_2 \partial_ z^2x_{2}(z,t) + \sqrt{|x_1(z,t)|} , \quad \forall (z,t) \in (0,\pi)\times \R_+,}\\
{x_2(0,t) = x_2(\pi,t)=0, \quad \forall t \in  \R_+.}
\end {array}
\right.
\end{equation}
Here $q_1,q_2>0$ are diffusion coefficients. 
This system can model a chemical reaction network with two reactants whose densities at a point $z$, and at time $t$ are given by $x_i(z,t)$, $i=1,2$.

We assume that $x_1 \in X_1:=L^2(0,\pi)$ and $x_2 \in X_2:=L^4(0,\pi)$. The state of the whole system \eqref{eq:Gekoppelte-parabolic-sys-ISS} is $X:=X_1 \times X_2$. 
Let us choose the following Lyapunov functions for subsystems 1 and 2 of \eqref{eq:Gekoppelte-parabolic-sys-ISS}, respectively:
\[
V_1(x_1)=\int_0^\pi{x_1^2(z)dz} = \|x_1\|^2_{L^2(0,\pi)}, \quad \forall x_1 \in  X_1,
\]
\[
V_2(x_2)=\int_0^\pi{x_2^4(z)dz} = \|x_2\|^4_{L^4(0,\pi)}, \quad \forall x_2 \in  X_2.
\]
Assume for a while that the maps $x_i$ are twice continuously differentiable.
Consider the Lie derivative of $V_1$:
\begin{eqnarray*}
\dot{V}_1(x_1) &=& 2 \int_0^\pi{x_1(z) \left( q_1  \partial_ z^2x_{1}(z) + x^2_2(z) \right)dz} \\
&\leq&
-2q_1 \left\| \partial_ z x_{1} \right\|^2_{L^2(0,\pi)} + 2\|x_1\|_{L^2(0,\pi)} \|x_2\|^2_{L^4(0,\pi)}.
\end{eqnarray*}
In the last estimate, we used integration by parts for the first term and the Cauchy-Schwarz inequality for the second one.

Applying Friedrichs' inequality to the first term, we obtain the estimate
\begin{eqnarray*}
\dot{V}_1(x_1) & \leq & -2q_1 \|x_1\|^2_{L^2(0,\pi)} + 2\|x_1\|_{L^2(0,\pi)} \|x_2\|^2_{L^4(0,\pi)} \\
&= &
-2q_1  V_1(x_1) + 2 \sqrt{V_1(x_1)} \sqrt{V_2(x_2)}.
\end{eqnarray*}
Take
\[
\chi_{12}(r)= \frac{1}{a} r,\; \forall r>0,
\]
with an arbitrary $a>0$. We obtain for all $x_1 \in X_1$, $x_2\in X_2$ that
\begin{eqnarray}
V_1(x_1) \geq \chi_{12}(V_2(x_2)) \qrq \frac{d}{dt}V_1(x_1) \leq
 -2(q_1-a^{\frac{1}{2}})  V_1(x_1).
\label{eq:V1dot_ineq}
\end{eqnarray}
The derivation was made under the assumption that
$x_1,x_2$ are twice continuously differentiable functions. The above estimate holds due to the density argument for general $x_1 \in L^2(0,\pi)$.
To ensure that $V_1$ is an ISS Lyapunov function, the following condition must hold:
\begin{eqnarray}
2(q_1-a^{\frac{1}{2}})>0 \quad\Iff\quad  a<q_1^2.
\label{eq:V1dot_ineq_koef}
\end{eqnarray}

Now assuming that $x_1$ and $x_2$ are smooth enough, consider the Lie derivative of $V_2$:
\begin{align*}
\dot{V}_2(x_2) &= 4 \int_0^\pi x_2^3(z)\Big( q_2 \partial_ z^2 x_{2}(z) + \sqrt{|x_1(z)|}\Big)dz \\
&\leq - 3q_2  \int_0^\pi 4 (\partial_ z x_{2})^2(z) x_{2}^2(z)  dz + 4 \int_0^\pi  x_2^3(z) |x_1(z)|^{\frac{1}{2}} dz\\
&= - 3q_2  \int_0^\pi \Big( \partial_z(x^2_2) \Big)^2 dz + 4 \int_0^\pi  x_2^3(z) |x_1(z)|^{\frac{1}{2}} dz.
\end{align*}

Applying Friedrichs' inequality to the first term (note that $x_2^2 \in L^2(0,\pi)$) and the H\"older's inequality to the last one, we obtain
\[
\dot{V}_2(x_2) \leq - 3q_2  V_2(x_2) +4 (V_2(x_2))^{3/4}(V_1(x_1))^{1/4}.
\]
Let
\vspace{-2mm}
\[
\chi_{21}(r)= \frac{1}{b} r,\; \forall r>0,
\]
where $b>0$ is an arbitrary constant. The following implication holds:
\[
V_2(x_2) \geq \chi_{21}(V_1(x_1)) \quad \Rightarrow \quad
\dot{V}_2(x_2) \leq  -(3q_2 - 4b^{\frac{1}{4}}) V_2(x_2).
\]
To ensure that $V_2$ is an ISS Lyapunov function for the second subsystem, the following condition must hold:
\begin{eqnarray}
3q_2 - 4b^{\frac{1}{4}}>0 \quad\Iff\quad  b<\Big(\frac{3q_2}{4}\Big)^4.
\label{eq:V2dot_ineq_koef}
\end{eqnarray}
To ensure the stability of the interconnection, we apply the
small-gain condition
\vspace{-1mm}
\begin{eqnarray}
\label{H_Function}
\chi_{12} \circ \chi_{21} < \id    \quad \Iff \quad ab>1.
\end{eqnarray}
If we impose the condition
\begin{eqnarray}
q_1^2\Big(\frac{3q_2}{4}\Big)^4 >1,
\label{eq:Conditions_on_q1_and_q2}
\end{eqnarray}
then it is possible to find $a,b$ so that the conditions \eqref{eq:V1dot_ineq_koef}, \eqref{eq:V2dot_ineq_koef}
and small-gain condition \eqref{H_Function} will be satisfied.
Thus, if $q_1$ and $q_2$ satisfy \eqref{eq:Conditions_on_q1_and_q2}, the small-gain theorem guarantees the UGAS property of the interconnection.

\section{Applications and further studies}

Applications of ISS for parabolic equations include many applications, such as the Stephan problem for two-phase dynamics (see \cite[Chapter 4]{koga2020materials}) for estimation and control. It also includes nuclear fusion control, as studied in \cite{AWP14}. 
More specifically, the following model for the dynamics of the safety factor profile $x$ in a tokamak is derived in  \cite{AWP14}, in polar coordinates
\begin{equation} \label{model}
\partial_ t x(r,t) = \partial_ r\left( \frac{\eta(r,t)}{r} \partial_ r\Big( r x(r,t)\Big) \right)+ \partial_ r\Big( \eta u(r,t)\Big), \quad  \forall (r,t) \in (0,1) \times \R_+,
\end{equation}
with Dirichlet boundary conditions:
\begin{equation}
x(0,t)= 
x(1,t)=0, \quad \forall t \in \R_+,
\label{BC}\end{equation}
and initial condition:
\begin{equation}\label{IC}
x(r,0)=x_{0}(r), \quad \forall  r \in (0,1),
\end{equation}
for a suitable function $x_0$.
Consider the following candidate Lyapunov function:
\begin{equation} \label{CLF}
V(x):=\frac{1}{2} \int_0^1 f(r)  x^2(r) dr, \quad x\in L^2((0,1),\R),
\end{equation}
where $f:[0,1]\rightarrow \R$ is a certain positive and twice continuously differentiable function. The next theorem is proven in \cite{APW13b}.
\begin{theorem} \label{thmISS} If there exist a positive function $f$, as introduced in \eqref{CLF}, and a positive constant $\alpha$ such that the following inequality is verified:
\begin{equation} \label{ineq}
f''(r)\eta(r,t) + f'(r) \left( \partial_r \eta(r,t) - \eta(r,t) \frac{1}{r} \right) +f(r) \left( \partial_r \eta(r,t)\frac{1}{r} -\eta(r,t) \frac{1}{r^2} \right) \leq -\alpha f(r), \quad \forall (r,t) \in (0,1)\times \R_+,
\end{equation}
then the following dissipation inequality holds along the solutions to \eqref{model}, \eqref{BC}, \eqref{IC}:
\begin{equation} \label{Vdotineq}
\dot{V}_u\leq -\alpha V\big(x(r,t)\big) +\int_0^1 f(r) \partial_r \Big( \eta(r,t) u \Big) x(r,t) dr, \quad \forall t \in \R_+.
\end{equation}
\end{theorem}

The previous result is a dissipative condition from which input-to-state stability properties have been derived in \cite{APW13b} when input disturbances are defined by actuation errors or estimation errors of the safety profile for the 
 PDE \eqref{model} with the boundary conditions \eqref{BC}. Moreover, it is used in \cite{APW13b} for the safety factor profile control for
the Tore Supra tokamak. Using this design, some experiments have been performed in \cite{mavkov2018experimental} for the TCV L-mode discharges.

ISS of hyperbolic equations has been studied for various applications, including traffic control as in \cite[Chap. 11.5]{yu2022traffic}. See also \cite{zhang2021boundary} where the Aw-Rascle-Zhang model for the macroscopic traffic flow dynamics is recalled, which reads for all $(r,t) \in [0,L]\times \R_+$ as
\begin{equation} 
\label{eqn1}
\left\{ {\begin{array}{*{20}{l}}
{{\partial _t}{\rho (r,t)} + {\partial _r}\left( {{\rho (r,t)}{v(r,t)}} \right) = 0},\\
{{\partial _t}\left( {{v}(r,t) + p\left( {{\rho(r,t)}} \right)} \right) +
{v}(r,t){\partial _r}\left( {{v}(r,t) + p\left( {{\rho (r,t)}} \right)}
\right) = \frac{{V\left( {{\rho (r,t)}} \right) - {v(r,t)}}}{\tau }},
\end{array}} \right.
\end{equation}
where $L$ is the length of the road segment, ${\rho}:[0,L]\times\R_+\rightarrow \R$ is the
vehicle density, ${v}:[0,L]\times\R_+\rightarrow \R$ is the
average speed, and $p\left( {{\rho }} \right)$ is an increasing
pressure function. The constant scalar $\tau $ is the relaxation term related to the
driving behavior, and 
$V\left( {{\rho}}
\right)$ is the speed-density fundamental diagram. See \cite[Chapter 1]{yu2022traffic} for more details on this model.
Equation \eqref{eqn1} is a system of two quasilinear balance laws for the traffic on a road segment. Considering boundary feedback controls and interconnecting the traffic equation \eqref{eqn1} on a freeway traffic network with $N$ segments yields a quasilinear hyperbolic system of size $2N$. Considering a suitable constant steady state of the flow on this network, linearizing the dynamics along this equilibrium, and using an appropriate change of variable denoted $x$, the following $2N$-order linear hyperbolic system of balance laws is derived in \cite{zhang2021boundary}:
\begin{equation} \label{eqn36}
{\partial _t}x(r,t)  + \Lambda {\partial_r}x(r,t)  = Mx(r,t)  + b,\quad \forall (r,t) \in [0,L]\times \R_+,
\end{equation}
where $
\Lambda$ is a diagonal invertible matrix in $\R^{2N\times 2N}$, $M$ is a matrix in $\R^{2N\times 2N}$ and $b$ is a constant bias vector in $\R^{2N}$. The boundary conditions are written as
\begin{equation} \label{eqn41}
{x _{in}}(t) = G{x _{out}}(t) + \theta(t),\quad \forall t\in \R_+,
\end{equation}
where $x _{in}$ and $x _{out}$ are respectively the input and the output of the system \eqref{eqn36} written in Riemann coordinates, $G$ is a control matrix and $\theta$ is a continuous time-varying disturbance with values in $\R^N$. 
Using a Lyapunov function candidate of \cite{BaC16}, the following result is obtained in \cite{zhang2021boundary}:
\begin{theorem}\label{thm:1}
Under suitable assumptions, the system (\ref{eqn36})  with the boundary feedback condition (\ref{eqn41}) ISS with respect to the inputs $b$ and $\theta(t)$. More specifically, there exist $\beta\in \mathcal{KL}$  and $\gamma\in \mathcal{K}_\infty$ such that, for any initial state ${x _0} \in {L^2}( (0,1),\R )$, for all $b\in\R^{2N}$, for all continuous function $\theta:\R_+\rightarrow \R^N$, the solution $x$ of (\ref{eqn36}) and (\ref{eqn41}) with $x(0)=x_0$ satisfies, for all $t\geq 0$ 
\begin{equation}
{{\left\| {x \left( { \cdot ,t} \right)} \right\|_{{L^2}( (0,1),\R )}} \leq \beta \left( {{{\left\| {{x _0}} \right\|}_{{L^2}( (0,1),\R )}}}, t\right)+ \gamma \left( {\mathop {\sup }\limits_{0 \le \tau  \le t} {{|{\left(\|b\|, \|\theta \left( \tau  \right)\| \right)} |}}} \right).}
\end{equation}
\end{theorem}
 Let us emphasize that the assumptions needed in the previous theorem are written in terms of  numerically tractable matrix inequalities. Thus numerical methods are obtained for the design of boundary controllers of traffic flow in a network, as done for experiments on the Fourth-Ring road of Beijing, China (see \cite{zhang2021boundary}).

\section{Conclusion and open problems}

This paper discusses the notion of input-to-state stability (ISS) for distributed parameter systems. We focus on the Lyapunov theory including characterizations of the ISS in terms of ISS Lyapunov functions, as well as Lyapunov-based small-gain theorems. We recall dissipative sufficient conditions for ISS of some classes of infinite-dimensional systems, such as those modelled by parabolic equations or hyperbolic systems.
We could not cover in this short introductory article the broad palette of the theoretic foundations and applications of the ISS theory. 
For a comprehensive overview of ISS including small-gain theory for analysis of networks, and for many more references, we refer to the survey \cite{MiP20}, and monographs \cite{KaK19,Mir23}. 
For the ISS theory of linear infinite-dimensional systems, we refer to \cite{Sch20}, for the ISS of delay systems (which are an important special case of the distributed parameter systems that we considered in this article), we refer to a recent survey \cite{CKP23}.

%

%

\begin{ack}[Acknowledgments]
The work of the second author is supported by MIAI Grenoble Alpes (ANR-19-P3IA-0003).
\end{ack}

\seealso{Refer to other chapters here (in particular the other chapter of A. Mironchenko and of C. Prieur)}

\bibliographystyle{Harvard}
\bibliography{Mir_LitList_NoMir,MyPublications}

\end{document}